\documentclass[12pt]{article}
\usepackage{graphicx}
\usepackage{verbatim}
\usepackage{harvard}

\renewcommand{\thefootnote}{\fnsymbol{footnote}}

\newcommand{\req}[1]{(\ref{#1})}
\newcommand{\be}{\begin{equation}}
\newcommand{\ee}{\end{equation}}
\newcommand{\bea}{\begin{eqnarray}}
\newcommand{\eea}{\end{eqnarray}}

\newcommand{\avg}[1]{\langle{#1}\rangle}

\newcommand{\BE}{\begin{eqnarray}}
\newcommand{\EE}{\end{eqnarray}}
\newcommand{\BEn}{\begin{eqnarray*}}
\newcommand{\EEn}{\end{eqnarray*}}
\newcommand{\barr}{\begin{array}}
\newcommand{\earr}{\end{array}}

\newcommand{\bit}{\begin{itemize}}      
\newcommand{\eit}{\end{itemize}}
\newcommand{\bc}{\begin{center}}
\newcommand{\ec}{\end{center}}
\newcommand{\ben}{\begin{enumerate}}    
\newcommand{\een}{\end{enumerate}}

\newcommand{\eps}{\epsilon}

\begin{document}

\title{Inter-pattern speculation: \\beyond minority, majority and \$-games}
\author{Damien Challet\\
  Oxford University\footnote{Nomura Centre for Quantitative Finance Mathematical Institute, Oxford University,  24--29 St Giles', Oxford OX1 3LB,  United Kingdom. Phone: +44 1865 280608, email: challet@maths.ox.ac.uk. I thank T. Bochud, S. Howison, W. Shaw and Y.-C. Zhang for their useful comments, and Wadham College for support.}. }

\date{\today}
\maketitle

\begin{abstract}
	A new simple model of financial market is proposed, based on the sequential and inter-temporal nature of trader-trader interaction, and on a new simple trading strategy space. In this pattern-based speculation model, the traders open and close their positions explicitly. Information ecology is strikingly similar to that of the Minority Game which suggest to reinterpret the latter as a model of synchronisation of predictability exploitation. Naive and sophisticated agents are shown to give rise to very different phenomenology \end{abstract}
\bigskip

{\bf JEL} G14, C73

{\em keywords:} agent-based modelling, speculation, efficiency, minority game


\newpage
\renewcommand{\thefootnote}{\arabic{footnote}}
\setcounter{footnote}{0}
Agent-based modelling is a way to mimic financial market phenomenology that focuses on individual behaviour \cite{SantaFeMarket,LLSS,Lux,BH97,Hommes_HAM_review,CZ97,BG01,dollargame,Cont2}. 
Because this approach requires to find the right  balance between simple and non-trivial rules, the complexity of the agents varies tremendously from model to model. At the same time, such models bring insights into financial market dynamics as long as they are amenable to analysis. As a consequence, the best strategy for building and studying this kind of models is to start from well-understood specimens, and either to push their boundaries, or to borrow their framework and create a new breed of models.

The best analysis is provided by exact solutions. However, heterogeneous agent-based models (HAM) are inherently harder to solve analytically than traditional representative agent models (see however \citeasnoun{Hommes_HAM_review} for a review of the Economics and Behavioural Finance literature on tractable HAM). Parameter heterogeneity, for instance in the case of price return extrapolation of trend-followers, can be solved by standard methods (see \citeasnoun{BH97} and variants). But for more complex heterogeneity, as when the pair-wise interaction strength varies amongst the agents, more powerful methods are needed. This is why the solution to the El Farol bar problem \cite{Arthur} could not be found until sophisticated methods first designed for disordered physical systems \cite{MPV} were used \cite{CMO03,CoolenBook}; in principle, these methods solve exactly any HAM where the interaction strength is independent from the physical distance between two agents. This is why HAM are a natural meeting point of Economics and Physics. An important contribution of Econophysics to HAM is the derivation of the Minority Game (MG) \cite{CZ97} from the El Farol bar problem,  and its numerous variants \cite{web}. Players in a minority game are rewarded when they act against the crowd. One key ingredient of the MG is a simple, well-defined strategy space of finite dimension, made up of look-up tables prescribing a binary action for all possible market states.

Most of HAM literature, both from Economics and Physics, rests on the assumption that the agents have an action horizon of not more than one time step. In other words, they take a decision at a given time step in order to optimise some quantity at next time step. This might be justified if the considered time step is long enough.  However it is obvious that real-life speculation implies that the traders have partially overlapping asset holding periods, with no systematic synchronisation. The closest approach in Economics literature is that of overlapping generation models and their numerous extensions  \cite{AllaisOverlappingGenerations,SamuelsonOverlappingGenerations,GeneakoplosReviewOLG}. The very fact that after many years almost all agent-based market models are still built with one-time step strategies\footnote{In some models the agents do try to predict two time-steps ahead, but cannot hold their positions \cite{Pawel}.}  is an indication that finding a strategy space of finite dimension suitable for the proper modelling of speculation is difficult. 
The challenge is to find a strategy space that allows for inter-temporal actions, hence, that should give a good reason to the traders for opening and closing their positions, and an equally good reason for holding them. Its parametrisation should be as simple as possible so as to minimise the number of parameters, while having a dimensionality under control.

Although the MG is sometimes regarded as having some relationship with financial markets, the latter is intriguing and has been difficult to understand fully. On the one hand, a minority game is compatible with the requirements that markets are competitive and negative sum games from the point of view of investors. On the other hand bubbles do not exist in Minority Games, and seem to be a consequence of a kind of majority game. Markets are rather escape games \cite{SZ99}, where one wishes to enter the market before the majority enters and to withdraw before the majority withdraws, i.e. anticipate to the crowd twice; \$-games  reduce the escape game by only asking to anticipate the majority of the next time step \cite{BG01,dollargame}.

In this paper, I first derive the correct market payoff that clarifies once for all the minority/majority nature of financial markets and then introduce a model of speculation based on a new simple and powerful strategy space. It will allow for the reinterpretation of the MG as a model of competition for predictability.



\section{Nature of financial markets}

The main reason why financial markets' nature is still somewhat unclear after so many years is due to their inter-temporal essence: all the traders are not active at the same time, hence perfect synchronisation cannot be achieved; even worse, all the trade orders arrive sequentially, hence simultaneity is a theoretician's phantasm at best ---which is rarely recognised in the literature. Let us review carefully the process of opening and closing a position.\footnote{The discussion below extends previous work \cite{CG04} to a finer time scale.}

Placing oneself in event time $\tau$, which increases by one unit whenever an order is placed, makes the discussion easier. Physical time will be denoted by $t$. At time $t_i=t(\tau_i)$ in physical time, or $\tau_i$ in event time, trader $i$ decides to open a position $a_i=\pm 1$ by placing a market order; at this time, the last paid price is $p(\tau_i)$.
\footnote{It is an order to buy/sell immediately at the best price. More patient traders place limit orders at or beyond best prices, thus obtaining a better deal.} The key observation is that his physical reaction time $\delta t_i$ that includes communication delays and possibly the time needed to make a conscious decision, results in a delay in event time of $\delta \tau_i$; $\delta \tau_i-1$ is the number of trades executed between $t_i$ and the transaction of agent $i$, hence, varies from transaction to transaction depending on market activity; the transaction takes place at time $\tau_i+\delta \tau_i$ and log-price $p(\tau_i+\delta \tau_i)$. 
\footnote{The smallest reaction time is around 1s for the DAX. J.-P. Bouchaud, private discussion.} 

\begin{figure}
\centerline{\includegraphics*[width=0.7\textwidth]{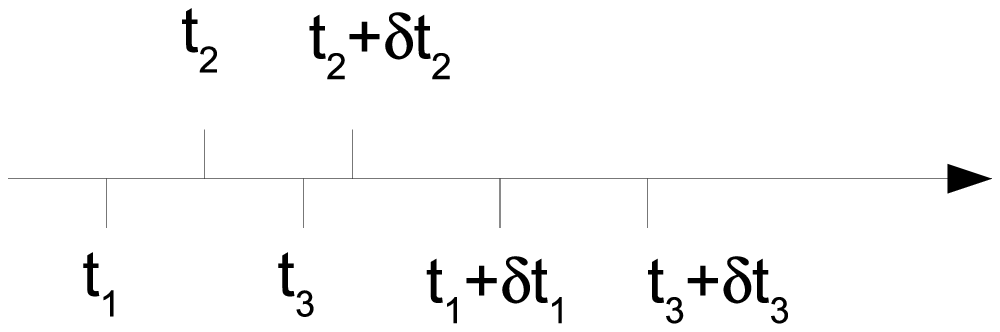}}
\caption{Time axis showing the intertwined nature of transactions caused by physical delays}
\label{fig:timeaxis}
\end{figure}

[Figure \ref{fig:timeaxis} around here]

During that time period, the order book changes; as a consequence $p(t_i+\delta t_i)$ differs from $p(t_i)$ by the sum of all the price changes caused by the $\delta\tau_i-1$ orders executed between $t_i$ and $t_i+\delta t_i$ that were placed by other agents $j$, $k$,\dots at time $t_j$, $t_k$,\dots and executed at time $t_j+\delta t_j$, $t_k+\delta t_k$,\dots (see Fig. \ref{fig:timeaxis}):
\be\label{Atj}
p(\tau_i+\delta \tau_i)=p(\tau_i)+\sum_{\hat\tau=\tau_i+1}^{\tau_i+\delta \tau_i}I[a(\hat\tau),\hat\tau]
\ee
where $I$ is the price impact function, that is, the price change due to a market or limit order, and $a(\tau')$ is the sign and size of the order placed at time $\tau'$. The presence of $\tau$ as a variable of $I$ allows for $a(\tau')$ being a limit order that modifies $I$. For the sake of simplicity, only market orders and linear impact $I(x)=x$ \cite{FarmerForce} will be considered here. In that case, 
\be 
p(\tau_i+\delta \tau_i)=p(\tau_i)+A(\tau_i,\delta\tau_i)
\ee
where $A$ is the accumulated price impact between $\tau_i+1$ and $\tau_i+\delta\tau_i$
\be\label{Ati}
A(\tau_i,\delta\tau_i)=\sum_{\tilde\tau=1}^{\delta\tau_i}a(\tau_i+\tilde\tau)=\sum_{\tilde\tau=1}^{\delta \tau_i}\sum_{j} a_j(\tau_j) \delta_{\tau_i+\tilde\tau,\tau_j+\delta\tau_j},
\ee
where the sum $\sum_j$ is over all the traders possibly interested in trading this particular asset and selects the agents whose transactions were executed between $\tau_i$ and $\tau_i+\delta\tau_i$. Equation \req{Ati} means that the group of traders with which trader $i$ interacts through market impact is different at each transaction, and that among this group, everyone has a different but partially common interacting group.

The position is held for some time. At $\tau'_i$, the trader decides to close his position, obtaining for the same reason as before $p(\tau'_i+\delta \tau'_i)$. His real payoff is 
\bea\label{marketmechanism}
a_i[p(\tau'_i+\delta \tau'_i)-p(\tau_i+\delta \tau_i)]&=& -a_iA(\tau_i,\delta\tau_i)\nonumber \\&&+a_i\sum_{\tau_i+\delta\tau_i<\tau\le\tau'_i} a(\tau) \nonumber\\&& -(-a_i)A(\tau'_i,\delta\tau_i').
\eea
The first and the last terms come from market impact and reaction time. They are  minority-game payoffs, which can be recognised at once by their `$-$' sign: one's market impact is reduced if one acts against the crowd (in this case, this means taking an action opposite to the majority of orders executed during the time delay). The central term, which has a `$+$' sign, represents the capital gain that could be achieved without reaction time nor market impact. It describes a {\em delayed} majority game, that is, a majority game to which trader $i$ {\em does not take part}: whereas $A(\tau_i,\delta_i)$ and $A(\tau'_i,\delta\tau_i')$ contain a contribution of trader $i$, the middle term does not.

The nature of financial markets depends therefore on the trading frequency and reaction time of each trader, and on the activity of the market: the relative importance of minority games decreases as the holding time $\tau'_i+\delta\tau_i'-\tau_i-\delta\tau_i$ increases; reversely, somebody who trades at each time step plays a minority game which is therefore the only consistent payoff in such a model where the time is not a coarse-grained quantity (e.g. a day). Interestingly, this is consistent with the behaviour of market makers who try and stabilise the price so as to minimise inventory risk, thereby inducing a mean-reverting behaviour, as in the minority game.

The case of a limit order is not more complicated: if at time $\tau_i$ agent $i$ places a limit order at $-a_i\Delta_i$ ($\Delta_i>0$) from the best price with timeout $T_i$, he expects that the majority of both price and limit orders will opposite to his, thereby obtaining a better price. Thus such a person is not only more patient, but also plays a minority game and fixes his reward. If his limit order is not executed before its timeout, the trader may place a new limit order at the same distance from the best price, which may have changed, and wait again. In this case, the price movement during the period where his first order was not executed is akin to increase $\delta t_i$ by $T_i$.


At this stage of the discussion a comparison between this market mechanism and the \$-game is in order. In the \$-game, a trader is rewarded if he buys a share at time $t'$, whose price increases at next time step $t'+1$: the payoff of trader $i$ is $a_i(t')A(t'+1)$, $t'$ being a physical discrete time that can be seen as a coarsening of $t$ or $\tau$; therefore, a \$-game also contains a delayed majority game. Replacing $t'+1$ by $\tau_i+\delta \tau$ and  assuming that trader $i$ opens a position at time $\tau$ and closes it at time $\tau+\delta \tau$, makes the \$-game payoff look like Eq. \req{marketmechanism} without market impact or reaction time, i.e. without minority games. There are three possible ways for that kind of payoff to appear: i) one knows in advance one's exact impact $A(\tau_i)$, which seems implausible; ii) one's reaction time is negligible, which only happens for infrequently traded stocks and if the size of the market order is smaller than the volume available at the best price, provided that the best prices do not change during the submission of the order; iii) the holding time  $\delta t$ is very large, in which limit market impact becomes much smaller than the price return $\delta t$. The fundamental problem in all these cases is that in the \$-game, $A(t'+1)$ also contains the contribution of trader $i$ as it forces the traders to be active at each time step and does not make it possible for the agents to hold and close their positions explicitly; a \$-game payoff is therefore inconsistent. Note finally that if the agents have expectations on the nature of the market, that is, on the middle term of Eq. \req{marketmechanism}, the \$-game payoff reduces to one time step, $t'$, and is a minority payoff for contrarians ($-a(t')A(t')$) and a majority payoff ($a(t')A(t')$) for trend-followers \cite{M01}.

The above discussion clearly shows the need for a strategy space that allows for holding a position, and above that gives real reasons to open and close positions. The experience of the Minority Game and El Farol bar problem suggests that the size of the strategy space should not depend on $N$ in order to make the model amenable to analytical solution.

\section{A model of speculation}

Finding a simple though meaningful speculation strategy space is surprisingly easy when examining the reasons why real traders open and close a position. It is reasonable to assume that they base their decisions on signals or patterns, such as mispricing (over/undervaluation), technical analysis, crossing of moving averages, news, etc. How to close a position is a matter of more variations: one can assume a fixed-time horizon, stop gains, stop losses, etc. I assume that the traders are inactive unless they receive some signal because some known pattern arises; this is to be compared with the many models where the agents trade at each time step \cite{Lux,Hommes_HAM_review,CZ97}. Therefore, the agents hold their positions between patterns. All the kinds of information regarded as possibly relevant by all the traders form the ensemble of the patterns which is assumed to be countable and finite.

Each trader recognises only a few patterns, because he has access to or can analyse only a limited number of information sources, or because he does not regard the other ones as relevant; in the present model, a trader is assigned at random a small set of personal patterns which is kept fixed during the whole duration of the simulation. Every time one of his patterns arises, he decides whether to open/close a position according to his measure of the average return between all the pairs of patterns that he is able to recognise. This is precisely how people using crossings of moving averages behave: take the case of a trader comparing two exponential moving averages (EMA) EMA100 and EMA200, with memory of 100, respectively 200 days: such trader is inactive unless EMA100 and EMA200 cross; the two types of crossing define two signals, or patterns. For instance, a set of patterns could be the 20 possible crossings between EMAs with memory length of 10, 20, 50, 100 and 200 days.

 The hope and the role of a trader are to identify pairs of patterns such that the average price return between two patterns is larger than some benchmark, for instance a risk-free rate (neglecting risk for the sake of simplicity); in this sense the agents follow the past average trend {\em between two patterns}, which makes sense if the average return is significantly different from the risk-free rate. In contrast with many other models (e.g. \citeasnoun{Lux}), the agents do not switch between trend-followers/fundamentalist/noise traders during the course of the simulation.

Defining what `average return' precisely means leads to the well-known problem of measuring returns of trading strategies in a back-testing, i.e., without actually using them. This is due to market impact and results usually in worse-than-expected gains when a strategy is used. Estimating correctly one's market impact is therefore a crucial but impossible aspect of back-testing because of two types of delusions. The first, temporal inaccuracy, is due to the over/underestimation of reaction time. Self-impact on the other hand is the impact that a real trade has on the price and market, which is not present in data used for back-testing. Both  of them cause imprecision in estimating returns of strategies not being used and, accordingly, both are actively investigated by practitioners. In the following one will call naive the agents who cannot measure exactly the market impact, while sophisticated agents will be those who can.



\subsection{Definition}
The mathematical definition of the model is as follows: $N$ traders can submit buy orders
($+1$) or sell orders ($-1$) or just be inactive. They base their decisions on patterns, denoted by $\mu$, and taking values $1,\cdots,P$. Each trader $i=1,\cdots,N$ is able to recognise $S$ patterns $\mu_{i,1},\cdots,\mu_{i,S}$, drawn at random and uniformly from $\{1,\cdots,P\}$ before the simulations commence; he is active, i.e., may wish to buy or sell one share of stock, only when the current pattern $\mu(t)$ is in his pattern list, that is, $\mu(t)\in\{\mu_{i,1},\cdots,\mu_{i,S}\}$. The kind of position he might take ($a_i(t)\in\{0,\pm1\}$) is determined by the average price return between two consecutive occurrences of  patterns. The time unit is that of pattern change, i.e., at each time step, $\mu(t)$ changes and is unique; for the time being, $\mu(t)$ is drawn at random and uniformly from $1,\cdots,P$. The duration of a time step is assumed to be larger than the time needed to place an order. The order in which agents' actions arrive is disregarded. Therefore, at time $t$, the volume is $V(t)=\sum_i |a_i(t)|$ and the excess return
$A(t)=\sum_{i=1}^Na_i(t)$ results in a (log-) price change of
\be
p(t+1)=p(t)+A(t).
\ee
$p(t+1)$, not $p(t)$, is the price actually
obtained by the people who placed their order at time $t$. Note that if the order of agents' actions were taken into account, one would first assign a random rank to the orders (possibly influenced by agents' reaction time); then the price obtained by the $n$-th order would be $p(t)+A(t)n(t)/V(t)$; if the reaction of all the agents is the same, the average effect of order arrival is irrelevant on average. 

There are  several ways to compute returns between two patterns. Assume that $\mu(t)=\mu$ and that $t'>t$ is the first subsequent occurrence of pattern $\nu$: $p(t')-p(t)$ is the price difference between these two patterns that does not take into account price impact, whereas $p(t'+1)-p(t+1)$ is the realistic price difference. Agent $i$ perceives a cumulative price return  $U_{i,\mu\to\nu}$ between pattern $\mu$ and $\nu$ which evolves according to 
\be
U_{i,\mu\to\nu}(t'+1)=U_{i,\mu\to\nu}(t)\nonumber+p(t'+1)-p(t+1)-(1-|a_i(t)|)\zeta_i [A(t')-A(t)].
\ee
when pattern $\mu$ has appeared but not $\nu$ yet, and stays constant otherwise; $\zeta$ is the naivety factor:  agents have $\zeta_i=1$ and fail to take reaction time into account properly, while or sophisticated, agents have $\zeta_i=0$ and compute perfectly the average price return. Finally, the variable $|a_i(t)|$ states whether the agent has opened/closed a position at time $t$ ($|a_i(t)|=1$), or was inactive ($|a_i(t)|=0$). When an agent trades, he perceives perfectly the effect of his reaction time whatever $\zeta_i$. In practice, $\zeta_i\ne 0$, and can be of any sign and value. This is because estimating reaction time exactly is impossible: even if real traders are often acutely aware of  its importance, they always over or underestimate it.

An agent only opens position between two of his patterns $\mu$ and $\nu$ if the average price return between them is larger than $\eps>0$, that is, if $|U_{i,\mu\to\nu}|>\eps t_{\mu\to\nu}$ where $t_{\mu\to\nu}$ is the total number of time-steps elapsed between patterns $\mu$ and $\nu$.

A further specification is that an agent trades between his $E$ best pairs of patterns, where $E\le S(S-1)/2$ is his maximal exposure, as one cannot trade from $\mu$ to $\nu$ and from $\nu$ to $\mu$ at the same time. If $E<S(S-1)/2$, the agent aims at trading only between his most profitable patterns; in this sense, the present model could be called the {\em pattern game}. In the following, the simplest case $S=2$, $E=1$ is analysed. The dynamics becomes simpler: when $\mu(t)=\mu_{i,1}$, if $|U_{\mu_{i,1}\to\mu_{i,2}}|>\eps t_{\mu_{i,1}\to\mu_{i,2}}$, he buys one share ($U_{i,\mu_1\to\mu_2}(t)>0$) or short-sells one share ($U_{i,\mu_1\to\mu_2}(t)<0$)\footnote{Short-selling consists in selling shares that one does not own. Closing such a position consists in
  buying back the stock later at a hopefully lower price.}, and holds his position until $\mu(t)=\mu_{i,2}$. When an agent closes and opens the same kind of position, he simply keeps his position open. 

Thus the basic model has five parameters: $N$, $S$, $P$, $\eps$ and $E$ and has the following ingredients: the agents are adaptive and heterogeneous, they have limited cognition abilities, and they can be naive or sophisticated.

Relevant quantities include the variance of $A$ 
\be
\sigma^2=\frac{\avg{A^2}-\avg{A}^2}{P},
\ee
the predictability seen by the naive agents
\be
J=\frac{1}{P(P-1)}\sum_{\mu,\nu,\mu\ne\nu}\avg{A(t)|\mu\to\nu}^2,
\ee
where $\avg{A(t)|\mu\to\nu}$ stands for the average price return per time step  between the occurrence of $\mu$ at time $t$ and the next occurrence of $\nu$. $J$ measures predictability that the naive agents with $\zeta=0$ hope to exploit; it is in fact related to absolute value of the average increase of $U_{\mu\to\nu}$ of naive agents per time step. Another closely related quantity is 
\be
K=\frac{1}{P(P-1)}\sum_{\mu,\nu,\mu\ne\nu}\avg{A(t+1)|\mu\to\nu}^2
\ee
is the actual exploitable predictability which is relevant to the sophisticated agents. Finally, a measure of price bias conditional to market states is given by the average return conditional to a given pattern
\be
H=\frac{\sum_{\mu}\avg{A|\mu}^2}{P}.
\ee

\begin{figure}
\centerline{\includegraphics*[width=0.7\textwidth]{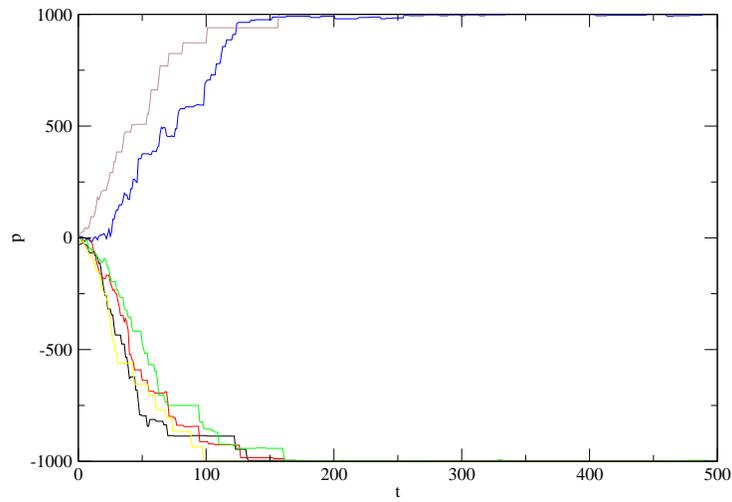}}
\caption{Price time series with $\eps=0$ ($N=1000$, $P=32$,  $S=2$, $\zeta=0$)}
\label{fig:S=2eps=0}
\end{figure}

[Figure \ref{fig:S=2eps=0} around here]

Before carrying out numerical simulations, one should keep in mind that the price is bounded between $-N$ and $+N$, since the traders are not allowed to have an exposure larger than 1. Assume that $\eps=0$, and that all the scores have  small random initial valuation (otherwise nobody trades in the first place). One observes in such case beginnings of bubbles or anti-bubbles, the price rising or decreasing to $\pm N$, and then staying constant. Indeed, the price increase/decrease is echoed in the scores of all the agents, which have all the same sign, therefore all stipulate the same action. The price is stuck (Fig \ref{fig:S=2eps=0}), as nobody wants to close his position, because everybody is convinced that the price should carry on on its way up/down. This phenomenon is found for all values of $\zeta$. 

\begin{figure}
\centerline{\includegraphics*[width=0.7\textwidth]{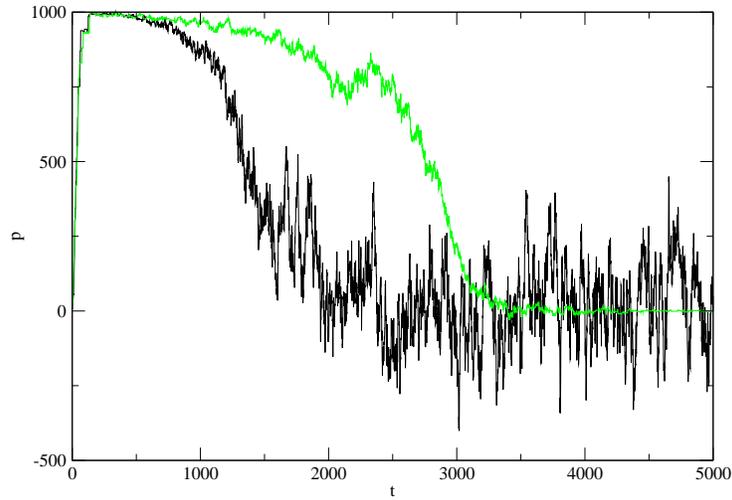}}
\caption{Price time series with $\eps=0.01$ ($N=1000$, $P=32$,  $S=2$). Black lines are for naive traders and green lines are for sophisticated traders.}
\label{fig:S=2eps>0}
\end{figure}

[Figure \ref{fig:S=2eps>0} around here]

\begin{figure}
\centerline{\includegraphics*[width=0.7\textwidth]{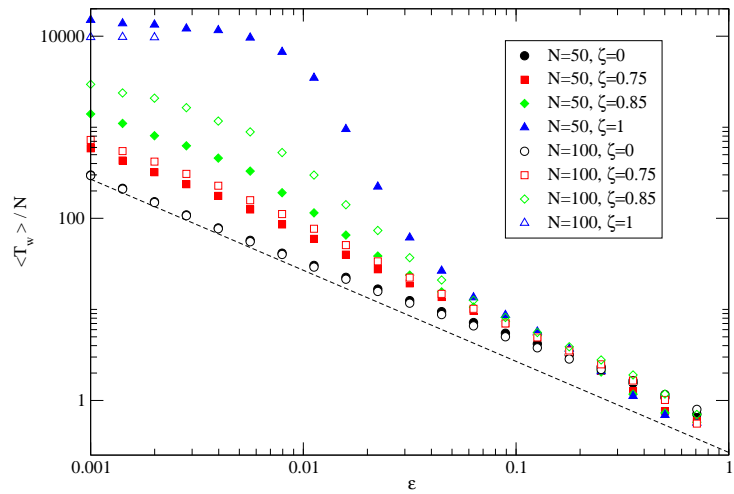}}
\caption{Scaled time to withdraw as a function of $\eps$. $P=10$, $N=50$ (full symbols) and $100$ (empty symbols), $\zeta=0$ (circles), $0.75$ (squares), $0.85$ (diamonds) and $1$ (triangles); average over 500 samples}
\label{fig:T/N}
\end{figure}

Figure \ref{fig:S=2eps>0} illustrates the typical price time series for $\eps>0$: first a bubble, then a waiting time until some traders begin to withdraw from the market. The price goes back to 0 and then fluctuates for a while. How these fluctuations are interpreted by the agents depends on $\zeta$ and $\eps$: increasing $\zeta$ makes it more difficult for the agents to understand that they should refrain from trading, because they are not capable of disentangling their own contribution from these fluctuations. Accordingly, the larger $\zeta$, the later the agents withdraw from the market, and the smaller $\eps$, the longer it takes (Fig. \ref{fig:T/N}). In this figure, the maximum number of iteration was capped at $10^6$; naive agents need therefore a very long time before withdrawing if $\eps$ is small. The scaling $T_w\propto N/\eps$ holds only for small $\zeta$ and $\eps$. For large $\eps$, $\zeta$ does not matter much.

[Figure \ref{fig:T/N} around here]

All the agents eventually withdraw from the market. This makes complete sense, as there is no reason to trade. Naivety results in a diminished ability to withdraw rapidly enough from a non-predictable market, and, as a by-product, in larger price fluctuations. This is consistent with the fact that volatility in real markets is much larger than if the traders were as rational as mainstream Economics theory assumes (see e.g. \citeasnoun{Shiller}). Naivety, an unavoidable deviation from rationality, is suggested here as one possible cause of excess volatility.
 

\subsection{Noise traders}

\begin{figure}
\centerline{\includegraphics*[width=0.7\textwidth]{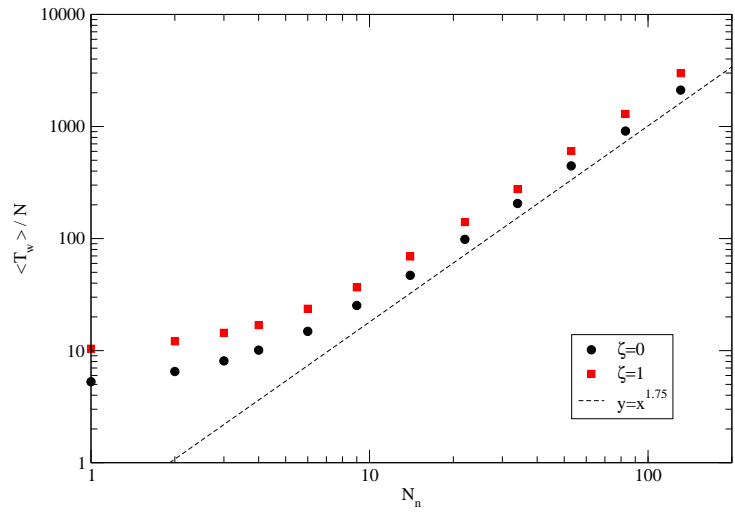}}
\caption{Scaled time to withdraw as a function of $N_n$. $P=10$, $N=100$, $\eps=0.1$, $\zeta=0$ (circles) and $1$ (squares); average over 500 samples}
\label{fig:T/Nnoise}
\end{figure}

[Figure \ref{fig:T/Nnoise} around here]

As the traders try and measure average returns, adding noise to the price  evolution ($A(t)\to A(t)+N_n\eta(t)$) where $\eta(t)$ is uniformly distributed between $-1$ and $1$) does not provide any trading opportunity, but makes it more difficult to estimate precisely average returns. Accordingly, the agents should withdraw later, and the larger $\zeta$, the later. This is precisely what happens: Fig \ref{fig:T/Nnoise} reports the average behaviour of $T_w$ for sophisticated and naive agents. One can therefore reinterpret this figure as an additional clue that naive agents are blinded by the fluctuations that they produce themselves.

\subsection{Market impact heterogeneity}



Real traders are heterogeneous in more than one way. In a population where each trader has his own $\zeta_i$, people with a small $\zeta_i$ evaluate gain opportunities better. This also means that the impact of people with large $\zeta_i$ provides predictability to the agents with a lower $\zeta_i$, and therefore the former are exploited by the latter,  giving a good reason to trade to sophisticated agents as long as naive agents are active.

\subsection{Market information structure}

Up to now, the model showed how hard it is not to trade. But how hard is it to make the agents want to trade? The problem is that they try to detect and exploit predictability, but that there is none when all the agents have the same abilities (e.g. the same $\zeta_i$). They might be fooled for a while by price fluctuations, as they try to detect trends between patterns, but eventually realise that there is no reason to trade. This is why the heterogeneity of the agents is crucial in this model. For instance, sophisticated agents would continue to trade if new naive agents replaced those who had understood their mistake. This would however probably preclude any hope to solve the model exactly. This is why I shall assume a different kind of heterogeneity. I will assume that there are people, the producers, who need to trade for other reasons than mere speculation \cite{ZMEM,MMM}: they use the market but make their living outside of it. Doing so they do not withdraw from the market, inject predictability and are exploited by the speculators. A simple way to include $N_p$ producers in the model is to assume that they induce a bias in the excess demand that depends on $\mu(t)$, i.e., $A(t)=A_{\rm prod}^{\mu(t)}+A_{\rm spec}(t)$. Each producer has a fixed contribution $\pm 1$ to $A_{\rm prod}^\mu$, drawn at random and equiprobably from $\{-1,+1\}$ for each $\mu$ and each producer, hence $A_{\rm prod}^\mu\propto \sqrt{N_p}$. In that way, the amount of market impact predictability introduced in the model is well controlled.

If there is no information structure in the market, i.e., if $\mu(t)$ does not depend at all on past patterns or price history, the effect of producers is akin to that of noise traders, hence, the speculators cannot exploit the predictability left by the producers. This is because the speculators need temporal correlations between occurring biases in order to exploit them, which happens when the transitions between market states are not equiprobable, i.e., when the transition matrix between patterns $W$ is such that $W_{\mu\to\nu}\ne 1/P$. This assumption is supported by empirical results: \citeasnoun{MarsiliClust}  determined states of the market with a clustering method, and found that the transitions between the states is highly non-random and has long-term memory which one neglects. Numerically, one chose to fix $W_{\mu\to\nu}$ to a random number between $0$ and $1$ and then normalised the transition probabilities; the variance of $W_{\mu\to\nu}$ is a parameter of the model which controls the amount of correlation between biases induced by the producers. It should be noted that the speculators are not able remove the global price bias because of the linearity of the price impact function. As a consequence, one should ensure that the effective average price return bias introduced by the producers is zero; it is nothing else but the sum of the components of $A_{\rm prod}$ weighted by the frequency of appearance of $\mu$, which is obtained from $W^\infty$.

\begin{figure}
\centerline{\includegraphics*[width=0.7\textwidth]{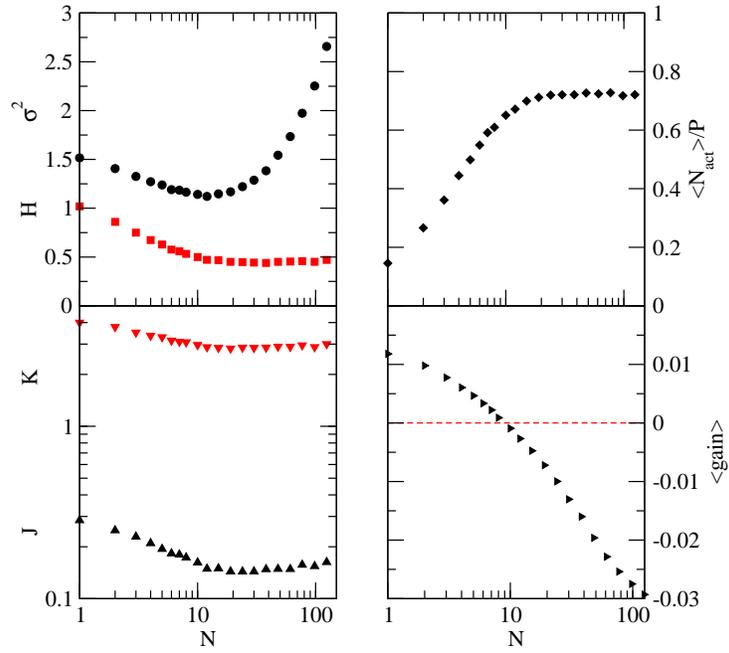}}
\caption{Volatility $\sigma^2$ (circles), price impact $H/P$ (squares), naive predictability $J$ (down triangles) and sophisticated predictability $K$ (up triangles), scaled fraction of active speculators (diamonds), and average gain per speculator and per time step (right triangles); $P=5$, $N_p=10$, $\eps=0.05$; average over 10000 samples of speculators}
\label{fig:ecology-4figs}
\end{figure}

[Figure \ref{fig:ecology-4figs} around here]

Adding more producers, that is, more predictable market impact, increases linearly the values of inter-pattern price return predictability measures  $J$ and $K$, as well as the average price impact $H$ and the fluctuations of the price $\sigma^2$. Then, keeping fixed the number of producers and their actions, adding more and more sophisticated speculators first decreases the fluctuations $\sigma^2$, the price impact $H$, and the exploitable predictability measures $J$ and $K$ (Fig. \ref{fig:ecology-4figs}); then all these quantities reach a minimum at an $N$ of order $P^2$, that is, of the order of the number of pairs of patterns. At the same time, the average total {\em number} of speculators in the market, denoted by $\avg{N_{\rm act}}$, reaches a plateau, thus the speculators are able to refrain from trading when predictability falls below some value determined by $\eps$. The fact that the fluctuations $\sigma^2$ increase for $N>P(P-1)/2$ means that the agents with the same pair of patterns enter and withdraw from the market in a synchronous way, leaving $\avg{N_{\rm act}}$, $H$, $J$ and $K$ constant. Avoiding such synchronicity can be realised by letting the agents have heterogeneous $\eps_i$ or $\zeta_i$. The average gain of the speculators is positive when their number is small enough but becomes negative for at $N\simeq P(P-1)/2$ in Fig \ref{fig:ecology-4figs}; if evolving capitals were implemented, less speculators would survive, which would lower the effective $N$, thus keeping the average gain at or above zero. This would also reduce or suppress the increase of $\sigma^2$.

The relationship between the producers and the speculators can be described as a symbiosis: without producers, the speculators do not trade; without speculators, the producers lose more on average, as the speculators reduce $H$. This picture, remarkably similar to that of the Minority Game with producers \cite{MMM,CCMZ,ZhangEcology}, justifies fully in retrospect the study of information ecology with minority games. 

\begin{figure}
\centerline{\includegraphics*[width=0.7\textwidth]{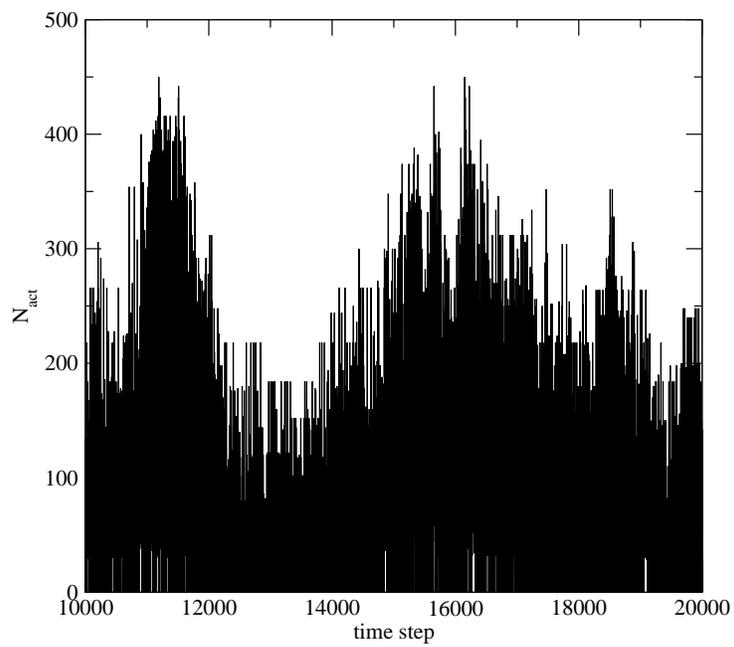}}
\caption{Number of active speculators as a function of time ($P=10$, $N=1000$, $N_p=100000$, $\eps=0.001$)}
\label{fig:Nact}
\end{figure}

[Figure \ref{fig:Nact} around here]

Guided by the knowledge that any mechanism subordinated to a random walk  and responsible for making the agents switch between being in and out of the market produces volatility clustering \cite{BouchMG}, one expects the emergence of such property in this model when the number of agents is large and $\eps>0$. This is exactly what happens, as shown by Fig. \ref{fig:Nact}, where the volume $N_{\rm act}$ displays a long term memory. Whether this model is able to reproduce faithfully real-market phenomenology is under investigation.

The similarity of information ecology between this model and the MG is a clue that the MG has some relevance for financial market modelling, and suggests to reinterpret it. Generally speaking, a minority mechanism is found when agents compete for a limited resource, that is, when they try to determine by trial and error the (sometimes implicit) resource level, and synchronise their actions so that the demand equals the resource on average \cite{CMO03,C04}. As a consequence, an agent is tempted to exploit the resource more than his fair share, hoping that the other agents happen to take a smaller share of it. The limited resource in financial markets is {\em predictability} and indeed information ecology has proven to be one of the most fascinating and plausible insights of minority games into market dynamics \cite{MMM,CCMZ}. Instead of regarding $A(t)$ in the MG as the instantaneous excess demand, one should reinterpret it as the {\em deviation from unpredictability} $A=0$ at time $t$. The two actions can be for instance to exploit an inefficiency ($+1$) or to refrain from it ($-1$); $A$ in this context would measure how efficiently an inefficiency is being exploited. Then everything becomes clearer: the fact that  $A(t)$ display mean-reverting behaviour is not problematic any more as it is not a price return. It simply means when the agents tend to under-exploit some predictability, they are likely to exploit it more in the next time steps, and reversely. What price return correspond to a given $A$ is not specified by the MG, but herding in the information space (the MG) should translate into interesting dynamics of relevant financial quantities such as volume and price; for instance, dynamical correlations of $|A|$ in the MG probably correspond to dynamical correlations in the volume of transactions. Therefore, studying the building up of volatility feedback, long-range memory and fat-tailed $A$ still makes sense, but not in a view to model explicitly price returns.

\section{Conclusion}

This model provides a new simple yet remarkably rich market modelling framework. It is readily extendable, and many relevant modifications are to be studied. The similarity of information ecology between the present model and the MG is striking and allowed for reinterpretation of the MG as model of competition for predictability, and substantiated the use of the MG to study predictability ecology of financial markets. Whether the proposed model is exactly solvable is under scrutiny.

Source code available at \verb+www.maths.ox.ac.uk/~challet+.
\bibliographystyle{kluwer}
\bibliography{patterns}
\newpage

\listoffigures

\newpage

\newpage

\newpage
\newpage
\newpage
\newpage
\newpage
\end{document}